\documentclass[prl,twocolumn,preprintnumbers,amsmath,amssymb,superscriptaddress]{revtex4-1}
\usepackage{graphicx,epsf,bm,amsmath}
\usepackage{subfigure,xcolor}
\usepackage[resetlabels]{multibib}
\usepackage{dsfont}

\usepackage{CJK}
\usepackage{soul}
\usepackage{color}
\definecolor{green}{RGB}{130,221,92}
\definecolor{red}{RGB}{252,103,105}
\definecolor{blue}{RGB}{62,206,251}

\newcommand{\bs}{\boldsymbol}
\newcommand{\be}{\begin{equation}}
\newcommand{\ee}{\end{equation}}

\begin{document}

\title{Synthesizing the Quantum Spin Hall Phase for Ultracold Atoms in Bichromatic Chiral Optical Ladders
}
\author{En Guo Guan}
\affiliation{School of Physical Science and Technology, Soochow University,
Suzhou 215006, China}
\author{Gang Wang}
\affiliation{School of Physical Science and Technology, Soochow University,
Suzhou 215006, China}
\author{Jian Hua Jiang}
\affiliation{School of Physical Science and Technology, Soochow
University, Suzhou 215006, China}
\author{Jun Hu}
\affiliation{School of Physical Science and Technology, Soochow
University, Suzhou 215006, China}

\author{Ray Kuang Lee}
\affiliation{Institute of Photonics Technologies and Physics
Department, National Tsing-Hua University, Hsinchu 300, Taiwan}

\date{\today}

\begin{abstract}
Realizing the topological bands of helical states poses a challenge
in studying ultracold atomic gases. Motivated by the recent
experimental success in realizing chiral optical ladders, here we
present a scheme for synthesizing topological quantum matter,
especially the quantum spin Hall phase, in the chiral optical
ladders. More precisely, we first establish the synthetic
pseudo-spin-orbit coupling and Zeeman splitting in the chiral
ladders. We analyze the band structure of the ladders exposed to the
bichromatic optical potentials and report the existence of quantum
spin Hall phase. We further identify a rich phase diagram of the
bichromatic chiral ladders, illustrating that our proposal features a
large space of system parameters exhibiting a variety of quantum
phase transitions. Our scheme can be readily implemented in the
existing experimental systems and hence provides a new method to
engineer the sophisticated topological bands for cold atomic gases.

\end{abstract}


\maketitle

Identifying ways to search and probe topological optical lattices and thus
topological quantum matter is a quest of major relevance in ultracold
atoms~\cite{CooperRMP2019,ZhuAP2019}. Over the past decade, much progress has
been witnessed in realizing the topological phases in cold atomic systems,
ranging from the Su-Schrieffer-Heeger model~\cite{Atala:2013}, the Hofstadter
model~\cite{AidelsburgerPRL2013,MiyakePRL2013,AidelsburgerNP2015} to the
Haldane model~\cite{Jotzu:2014}. Important cases are the quantum spin Hall
insulators (QSHIs) that arise in spin-orbit coupled systems, where the
existence of helical edge states produces a spin current along the edge of a
strip. Realizing the QSHI for ultracold gases was suggested in early
theoretical works~\cite{goldmanprl2010}. The QSHI would be implemented by
engineering the hyperfine structure of the ultracold atoms to synthesize the
non-Abelian gauge to mimic spin-orbit coupling~\cite{goldmanprl2010}. On the
experimental side, the only realized examples are cases where independent
quantum Hall (QH) insulators of opposite Chern number have been paired to
form a single system~\cite{AidelsburgerPRL2013,KennedyPRL2013}. The system
thereby is protected by a $\mathbb{Z}$ topological index. Despite these
advance, optical lattices featuring QSHI with spin-orbit interaction have so
far been lacking to our knowledge.

Along this line, chiral ladder systems for ultracold atoms constitute one of
timely topics of engineering the topological quantum matter with synthetic
gauge and synthetic dimensions~\cite{Paredes}. They represent a simple yet
effective platform to study exotic quantum phases of ultracold atoms, given
that ultracold atoms in optical lattices naturally realize such a strip
geometry~\cite{AtalaNP2014}. Apart from the ladder structure in the real
dimension~\cite{AtalaNP2014}, the internal degrees of freedom of atoms such
as the hyperfine states~\cite{StuhlScience2015,ManciniScience2015} and clock
states~\cite{ReyPRL2016,Livi2016,YeNature2017}, and the external degrees of
freedom such as the momentum states~\cite{GadwaySciAdv2017,WangPRL2019} and
lattice orbitals~\cite{ShinPRL2018,ShinPRL2019}, can be exploited to
fabricate the chiral ladders in the synthetic dimensions. To date, an intense
theoretical and experimental investigation has revealed rich topological
features of the chiral ladders~\cite{Paredes}. For example, the chiral ladder
systems in synthetic dimension have led to visualization of the chiral edge
states related to the QH phase~\cite{StuhlScience2015,ManciniScience2015}.

Motivated by the recent experimental realizations of ultracold atomic ladders
immersed in synthetic gauge
potentials~\cite{AtalaNP2014,ManciniScience2015,StuhlScience2015}, here we
propose a chiral-ladder realization of the topological bands of QSHIs. To be
specific, we utilize the leg degree of freedom to synthesize the
pseudo-spin-orbit coupling and Zeeman splitting. By exposing it to a 1D
bichromatic optical potential along the direction of legs, we construct the
model of bichromatic chiral ladder. We analyze the energy band of this
bichromatic chiral ladder, revealing the quantum spin Hall phase. We further
highlight the emergence of distinct topological phases by varying the
parameters of the ladder systems. Our study is of direct experimental
relevance for laboratories where ultracold gases within a chiral ladder
geometry are
realized~\cite{AtalaNP2014,StuhlScience2015,ManciniScience2015,Livi2016,YeNature2017,GadwaySciAdv2017,WangPRL2019,ShinPRL2018,ShinPRL2019},
hence providing a realistic way to achieve topological quantum matter of
ultracold atoms.

\begin{figure}[t]
  \includegraphics[width=1.0\linewidth]{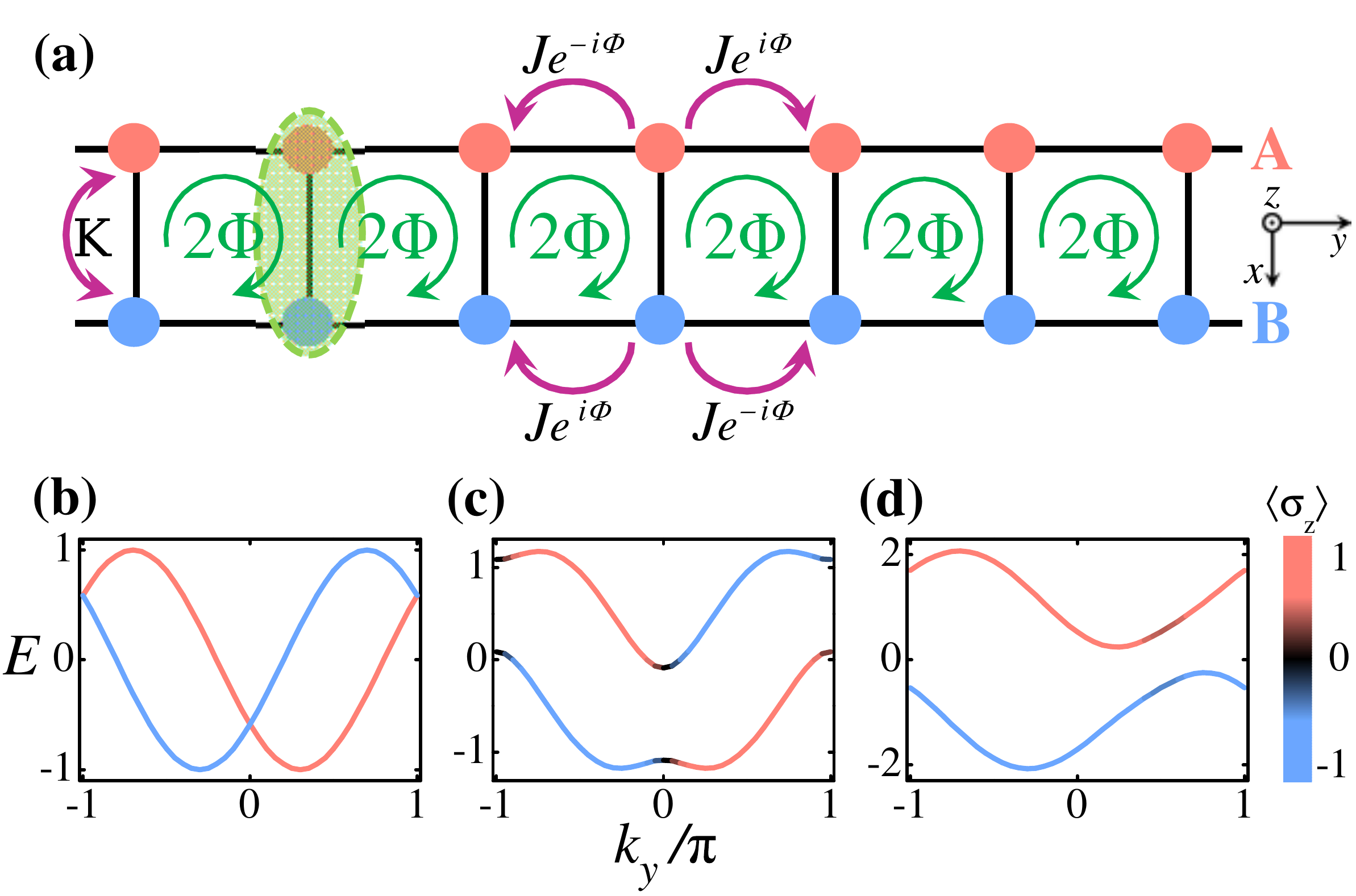}
  \caption{(a) Schematic representation of the two-leg ladder. This
optical lattice has a double-well structure in $x$ direction, while unlimitedly extends in
$y$ direction. $A$ and $B$ (the red and blue colors) label the two different
degrees of freedom of legs, denoting the two different pseudo-spin states.
$J$ and $K$ are the hopping
amplitudes along the legs and rungs, respectively. An artificial magnetic flux
$2\phi$ penetrates the plaquettes.
An additional potential difference $2\Delta$ is imposed between the legs.
(b)-(d) Band structures of the chiral ladders for various $K$'s and $\Delta$'s.
Energy is set in units of $J$. The color of the lines specifies the spin
magnetization of the Bloch state. The magnetic flux
is set as $2\phi=0.6\pi$. Other parameters: (b)$K=0,\; \Delta=0$ (c) $K=1,\;
\Delta=0$, (d) $K=1,\; \Delta=1$.
}\label{fig:schematic}
\end{figure}

Our starting point is the ladder geometry of optical lattices for
noninteracting \emph{spinless} fermions that has been synthesized
experimentally~\cite{AtalaNP2014}. The access of QSHE at first requires the
identification of two degrees of freedom representing the two spin states. We
resort to a two-leg ladder threaded by a uniform artificial magnetic
field~\cite{AtalaNP2014}, the so-called chiral ladder, to encode the two
degrees of freedom and enable their mutual interactions. As sketched in
Fig.~\ref{fig:schematic} (a), the chiral ladder consists of a two-leg strip
with intra- and inter-leg hoppings $J$ and $K$. Each plaquette encloses a net
gauge flux $2\phi$. According to the Peierls substitution, the Landau gauge
adopted here will imprint a phase factor $\pm i \phi$ on the hoppings along
the legs ($\pm$ for the A- and B-legs, respectively).
In addition, we superimpose an energy offset $2\Delta$ between the legs.
Physically, this corresponds to a deep double-well configuration oriented
along $x$ direction. The overall ladder Hamiltonian in real space reads:
\begin{align}
\mathcal{H} = &-J \underset{n}{\sum}\left(e^{i \phi} c^{\dagger}_{n+1,A} c^{}_{n,A}+e^{-i \phi}  c^{\dagger}_{n+1,B} c^{}_{n ,B} +\; h.c. \right) \nonumber \\
&+ \Delta \underset{n}{\sum} \left( c^{\dagger}_{n,A} c^{}_{n,A} -  c^{\dagger}_{n,B} c^{}_{n,B} \right) \nonumber \\
&-K \underset{n}{\sum} \left( c^{\dagger}_{n,A} c^{}_{n,B}\; +\;  c^{\dagger}_{n,B} c^{}_{n,A} \right). \label{eq:H}
\end{align}
Here the operator $ c^{\dagger}_{ n,\mu} \; \left( c^{}_{ n,\mu}\right)$
creates (annihilates) a fermionic particle on site $(n,\mu)$, where
$\mu=(A,B)$.

A precise connection can be made between a spin-orbit coupled chain and the
chiral ladder. Following the interpretation of Ref.~\cite{Huegel:2013}, one
can think of Eq.~(\ref{eq:H}) as a 1D optical lattice with pseudo-spins
represented by $A$- and $B$-legs. Keeping this in mind, we introduce the
spinor operator $\Psi^{\dagger}_{n}=\left(c^{\dagger}_{n,A}, \;
c^{\dagger}_{n,B}\right)^T$ and rearrange our model 
in the spinor space as follows:
\begin{eqnarray}
\mathcal{H} &=&\underset{n}{\sum} \Delta \Psi^{\dagger}_{n} \hat{\sigma}_z \Psi_{n}
-K \Psi^{\dagger}_{n} \hat{\sigma}_x \Psi_{n} \nonumber \\
&-&J \underset{n}{\sum} \Psi^{\dagger}_{n+1} e^{\imath \phi \hat{\sigma}_z} \Psi_{n}+h.c.,
\end{eqnarray}
where $\hat{\sigma}_{i}$ are the Pauli operators. Written in the
momentum-space the resulted Hamiltonian is of the form
\begin{equation}
\mathcal{H}=-\underset{k}{\sum} \Psi^{\dagger}_{k}M(k)\Psi_{k} \label{eq:chiralH}
\end{equation}
with $M(k)=2J \cos \phi \cos k \bs{\mathds{1}}+2J \sin \phi \sin k
\hat \sigma_{z} + K \hat \sigma_{x}-\Delta \hat \sigma_z.$ We
computed the the energy bands and the pseudo-magnetization $\langle
\sigma_z \rangle$ for a fixed flux $2\phi$ and different values of
inter-leg hopping $K$ and energy offset $2\Delta$, shown in
Fig.~\ref{fig:schematic}(b-d). Figure~\ref{fig:schematic}(b)
corresponds to the band structure in the case of vanishing $K$ and
$\Delta$. Clearly, one can observe a positive (negative) shift of
energy minimum for the pseudospin-up (-down) particles. This
evidences an effective spin-momentum locking derived from the nonzero
magnetic flux. Provided the inter-leg tunnelling $K$ is turned on, as
can be seen in Fig.~\ref{fig:schematic} (c), it opens a gap and the
states get increasingly spin-mixed. Thereby the $\hat \sigma_x$ term
in Eq.~\ref{eq:chiralH} brings about a spin-flip. Further, in the
case of large offset a spin separation is visualized from the spin
magnetization in Fig.~\ref{fig:schematic} (d). This term therefore
indicates a pseudo-Zeeman splitting. In total, the above proposed
chiral-ladder geometry can be mapped onto a 1D spin-orbit coupled
lattice with spin flip and Zeeman field. It is necessary to stress
that the pseudo-Zeeman term involved here is associated with the
double well, instead of the artificial magnetic field. This is
different from the electronic systems. As a result, it can give an
independent control over the ladder systems.

To realize the topological bands of chiral ladders, we propose to impose a
bichromatic optical superlattice $V(y)=V_p \cos^2(k_p y)+V_a
\cos^2(k_a+\theta)$ along $y$ direction. This can be created by superimposing
on a primary, deep lattice an auxiliary, weak lattice~\cite{RoatiNature2008}.
$V_p$ and $V_a$ separately denote the depth of the primary and auxiliary
lattice. $k_p$ and $k_a$ are the two lattice wave numbers. The variable phase
$\theta$ accounts for the relative position of the lattices. The modulation
is applied equally on the \textit{A}- and \textit{B}-legs (the shaded circle
of Fig.~\ref{fig:schematic} (a)). When the primary lattice depth is
sufficiently large, the ultracold atoms feel actually the onsite potential
energies as follows $V(n)=\Lambda \cos(2\pi \beta
n+\theta)$~\cite{RoatiNature2008}. Here $\Lambda$ characterizes the
modulation strength due to the auxiliary lattice, $\beta$ equals to the ratio
$k_a/k_p$. Summing up, the modulated optical ladder can described by a
tight-binding Hamiltonian
\begin{eqnarray}
\mathcal{H}_{1D}(n,\theta) = &-&\sum_{n} \Psi_n^{\dagger} \left[\Lambda\cos(2 \pi \beta n+\theta) \right] \Psi_{n} \nonumber\\
&+&\sum_{n} \Psi_n^{\dagger} \left[\Delta \hat{\sigma}_z - K \hat{\sigma}_x \right] \Psi_{n} \nonumber\\
&-& \sum_{n} J \Psi_{n+1}^{\dagger} e^{\imath \phi \hat{\sigma}_z} \Psi_n +
h.c..
\label{eq2:spinorbitcoupledHarper}
\end{eqnarray}
The first term describes the bichromatic modulation. In the second term
$\Delta$ and $K$ denote the strength of Zeeman splitting and pseudospin flip
specified above. And the last term contributes to the spin-orbit coupling. In
experiments, $\Lambda$ and $J$ can be controlled independently by varying the
depth of the primary and auxiliary lattice potentials. $\theta$ can be tuned
by the relative shift between the two lattices. $\Delta$ and $K$ can be
freely altered by tilting the double-well. This setting of optical lattices
enables us to continuously tune the system among various topological regimes,
as we elaborated below. Hereafter, the chiral ladders with the bichromatic
modulation will be referred to as ``bichromatic chiral ladders (BCLs)" for
convenience.

In order to identify the topological nature of the BCL structure, we
now establish the connection of the 1D BCL to a 2D spin-orbit coupled
square lattice pierced by a magnetic field. The 2D Hamiltonian can be
found by using the approach of dimensional
extension~\cite{ChenPRL2012,ZilberbergPRL2012,VerbinPRL2013,ZilberbergPRL2013,ZilberbergNPhys2016}.
To be specific, given the parameter $\theta$ is cyclically varied in
$[0,\; 2\pi]$, it can be regarded as a quasi-momentum $k_z$ along a
virtual coordinate $\hat{z}$. After we make substitutions of $J
\rightarrow t_{y}$, $\theta \rightarrow k_{z}$, and $\Lambda
\rightarrow 2 t_z$ and relabel the spinor as $\Psi^{\phantom
\dagger}_{n,k_z}$, the present 1D model can be converted into a 2D
Hamiltonian in a mixed momentum-position representation $\left(n, \;
k_z\right)$,
\begin{eqnarray}
\mathcal{H}_{2D}(n,k_z) = &-&\sum_{n,k_z} \Psi_{n,k_z}^{\dagger} \left[2t_z\cos(2 \pi \beta n+k_z)\right] \Psi_{n,k_z} \nonumber\\
&+& \sum_{n,k_z} \Psi_{n,k_z}^{\dagger} \left[\Delta \hat{\sigma}_z - K \hat{\sigma}_x \right] \Psi_{n,k_z} \nonumber \\
&-& \sum_{n,k_z} t_y \Psi_{n+1,k_z}^{\dagger} e^{\imath \phi \hat{\sigma}_z} \Psi_{n,k_z} + h.c.. \label{eq3:strip-spinorbitcoupled-H}
\end{eqnarray}
Performing the inverse Fourier transform, $\Psi^{\phantom \dagger}_{n,k_z} =
\sum_m e^{-i k_z m} \Psi^{\phantom \dagger}_{n,m}$, gives the real-space 2D
Hamiltonian
\begin{eqnarray}
\mathcal{H}_{2D}(n,m) = &-&t_z\sum_{n,m} e^{\imath 2\pi \beta n}
\Psi_{n,m+1}^{\dagger} \Psi_{n,m} + h.c. \nonumber\\
&-& t_y\sum_{n,m} e^{\imath \phi \hat{\sigma}_z}
\Psi_{n+1,m}^{\dagger} \Psi_{n,m} + h.c. \nonumber\\
&+& \sum_{n,m} \left[\Delta \hat{\sigma}_z - K \hat{\sigma}_x \right] \Psi_{n,m}^{\dagger}\Psi_{n,m}.
\label{eq4:2Dspinorbitcoupled-H}
\end{eqnarray}
This Hamiltonian exactly describes a spinor moving on a square
lattice defined in the $y-z$ plane $(y=n, z=m)$, which is threaded by a
uniform magnetic flux $\beta$ per plaquette. Based on the analogy
between our 1D system and the 2D system, this allows us to define the
topological properties of the BCL. That is the topological origin of
our model. We should emphasize that, different from the conventional
Harper-Hofstadter model~\cite{Harper,Hofstadter}, the spin-orbit
coupling, spin flip, and Zeeman splitting are involved in
Eq.~\ref{eq4:2Dspinorbitcoupled-H}.

\begin{figure}[htb]
\begin{center}
\includegraphics[width=\columnwidth]{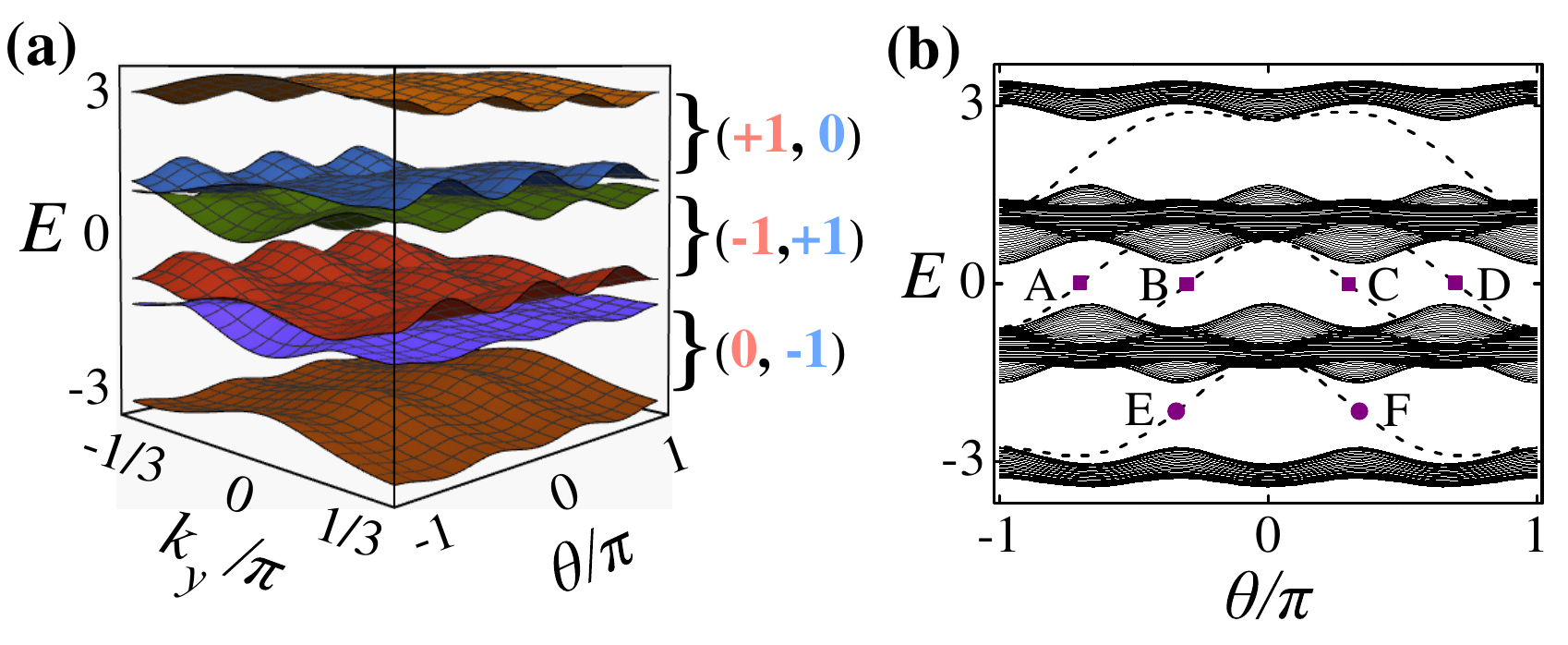}
\caption{Energy spectrum for $2\phi=0.6\pi$, $\Lambda=1.5$, $\beta=1/3$, $\Delta=1$, and $K=0.1$.
The energy is set in units of $J$. (a) Energy spectrum
$E=E (k_y, \theta)$ for the infinite BCL. The
integers near the graph label the spin Chern numbers of different bandgaps.
(b) Energy spectrum obtained from a finite ladder.
As a function of $\theta$, the spectrum is composed of the bulk bands (solid lines)
and dispersion curves that traverse the gaps (dashed lines).
$\blacksquare$ ({\large $\bullet$})
is the markers for the states at the same Fermi energy.
}
\label{fig:spectrum}
\end{center}
\end{figure}

Having precisely mapped the 1D BCL onto the analogous 2D system, we at this
point move to the topological phases of the BCLs. Herein, we assume a
rational $\beta$, i.e., $\beta=p/q$ with $p,\; q$ being coprime integers. The
length of unit cell of the ladder turns out to be $q$. Inserting the Bloch
waves $\Psi_{n+q}=e^{i k_y} \Psi_n$ in Eq.~\ref{eq2:spinorbitcoupledHarper}
yields the band structures $E=E (k_y, \theta)$ of the modulated ladder,
together with the associated eigenvectors, that are defined by $\theta$.
Figure~\ref{fig:spectrum} (a) displays the energy band for $\beta=1/3$.
To quantify the topological properties of our system, we calculate the
spin-up (spin-down, respectively) Chern numbers associated with the bandgaps
defined in the parameter space $(k_y, \theta)$. In Fig.~\ref{fig:spectrum}
(a) we have labeled the spin gap Chern numbers. We note that for the middle
gap the Chern numbers are given by $(C_{\uparrow}, \; C_{\downarrow})=(-1, \;
+1)$. This leads to the nontrivial $Z_2$ index $\frac{1}{2}
(C_{\uparrow}-C_{\downarrow})=-1$ which verifies the emergence of QSHI. It
should be noted that the time-reversal symmetry is broken in the analogous 2D
Hamiltonian, and thus this phase corresponds to the
time-reversal-symmetry-broken QSHI phase~\cite{ShengPRL2011}. In the
meanwhile, the Chern numbers of the first gap are $(C_{\uparrow}, \;
C_{\downarrow})=(0, \; -1)$. This indicates a spin-filtered QH
phase~\cite{GoldmanEPL2012,BeugelingPRB2012}. A similar result holds for the
third gap, but with $(C_{\uparrow}, \; C_{\downarrow})=(+1, \; 0)$.

\begin{figure}[htb]
\includegraphics[width=1.0\columnwidth]{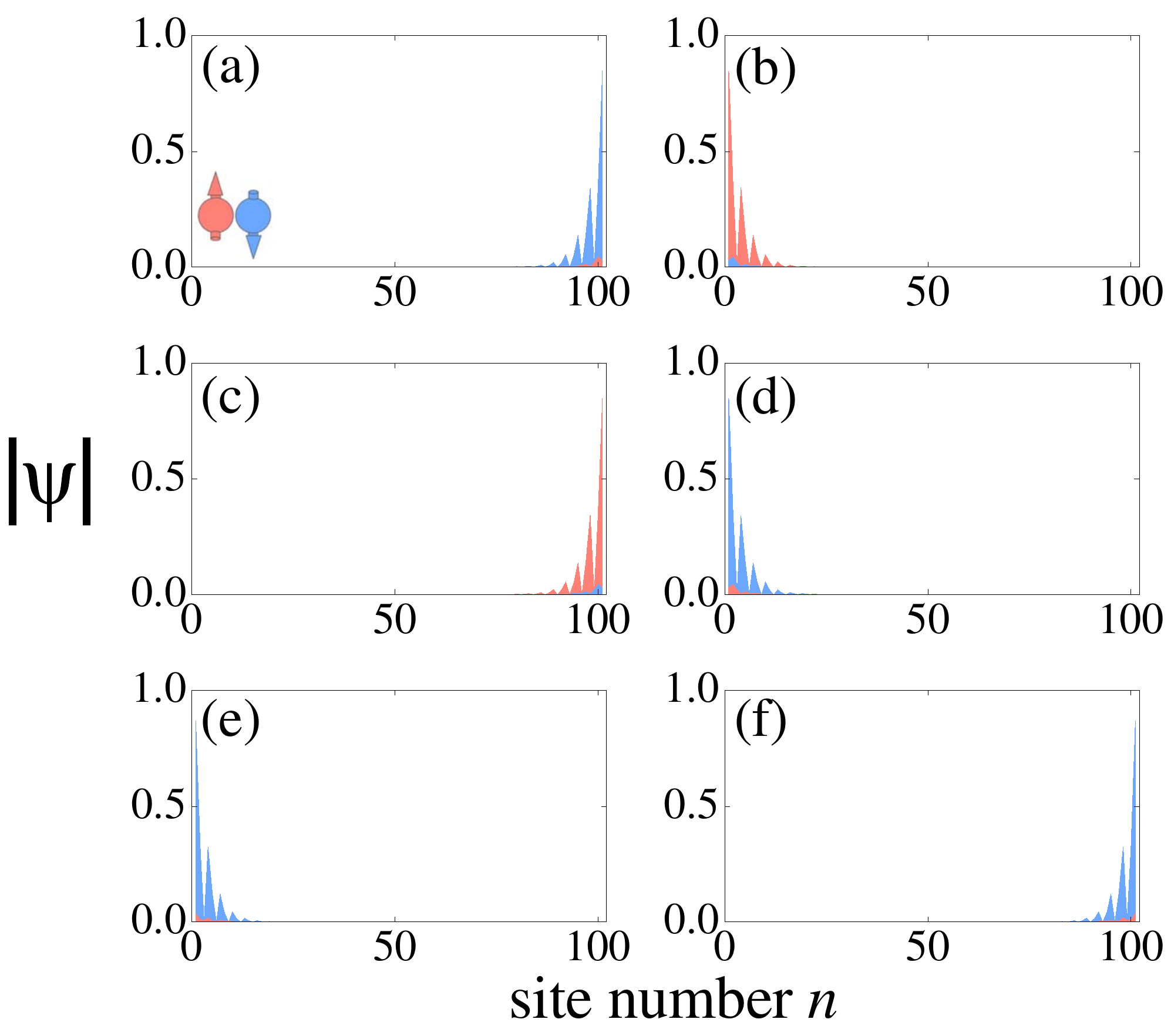}
\caption{The mode amplitudes of the gapless states in Fig.~\ref{fig:spectrum}(b).
(a-d) corresponds to the states marked by $\blacksquare$, while
(e,f) to the states {\large $\bullet$}. The spin component
$\Psi_{n\uparrow}$ (respectively, $\Psi_{n\downarrow}$) is
represented in red (respectively, blue).}
\label{fig:helicalchiral}
\end{figure}

The definite topological property in a finite system is the emergence of
gapless edge states as the phase $\theta$ varies. In
Fig.~\ref{fig:spectrum}(b), illustrating the spectrum of a finite ladder, the
dispersion curves of the additional states (dashed lines) are clearly
superimposed on the bulk bandgaps. At a given Fermi level
$E_{\text{Fermi}}=0$, gapless states, labeled as \textbf{A}, \textbf{B},
\textbf{C} and \textbf{D}, emerge in the middle bandgap. States \textbf{A}
and \textbf{C} are localized near $n=L$, while \textbf{B} and \textbf{D} are
localized near $n=0$ (Fig.~\ref{fig:helicalchiral} (a-d)). The slope of
dispersion curves in Fig.~\ref{fig:spectrum}(b) determines that the two
states \textbf{A} and \textbf{C} are counterpropagating in the analogous 2D
square lattice. Meanwhile, state \textbf{A} and \textbf{C} are almost fully
spin-down and spin-up polarized, respectively
(Fig.~\ref{fig:helicalchiral}(a), (c)). Therefore, these two pair of mid-gap
states form the helical edge states, indicating the QSHI phase of the middle
gaps. On the other hand, when the Fermi energy is adjusted inside the lowest
gap, two propagating states traverse the bulk gap (\textbf{E} and \textbf{F}
in Fig.~\ref{fig:spectrum} (b)). \textbf{E} and \textbf{F} are characterized
by the single spin-component which are localized at the opposite edges
(Fig.~\ref{fig:helicalchiral}(e,f)). In the analogous 2D system the
excitations \textbf{E} and \textbf{F} constitute a pair of chiral edges
states with single spin-component. This is associated with the spin-filtered
QH phase~\cite{GoldmanEPL2012,BeugelingPRB2012}. We emphasize that the
pseudospin components in our proposal are manifested as the leg degree of
freedom. As a result, these edge states are localized at the extremities of
the left- and right-legs, respectively. This will give convenience to the
direct observations of the topological phases.

\begin{figure}[htb]
\includegraphics[width=0.8\columnwidth]{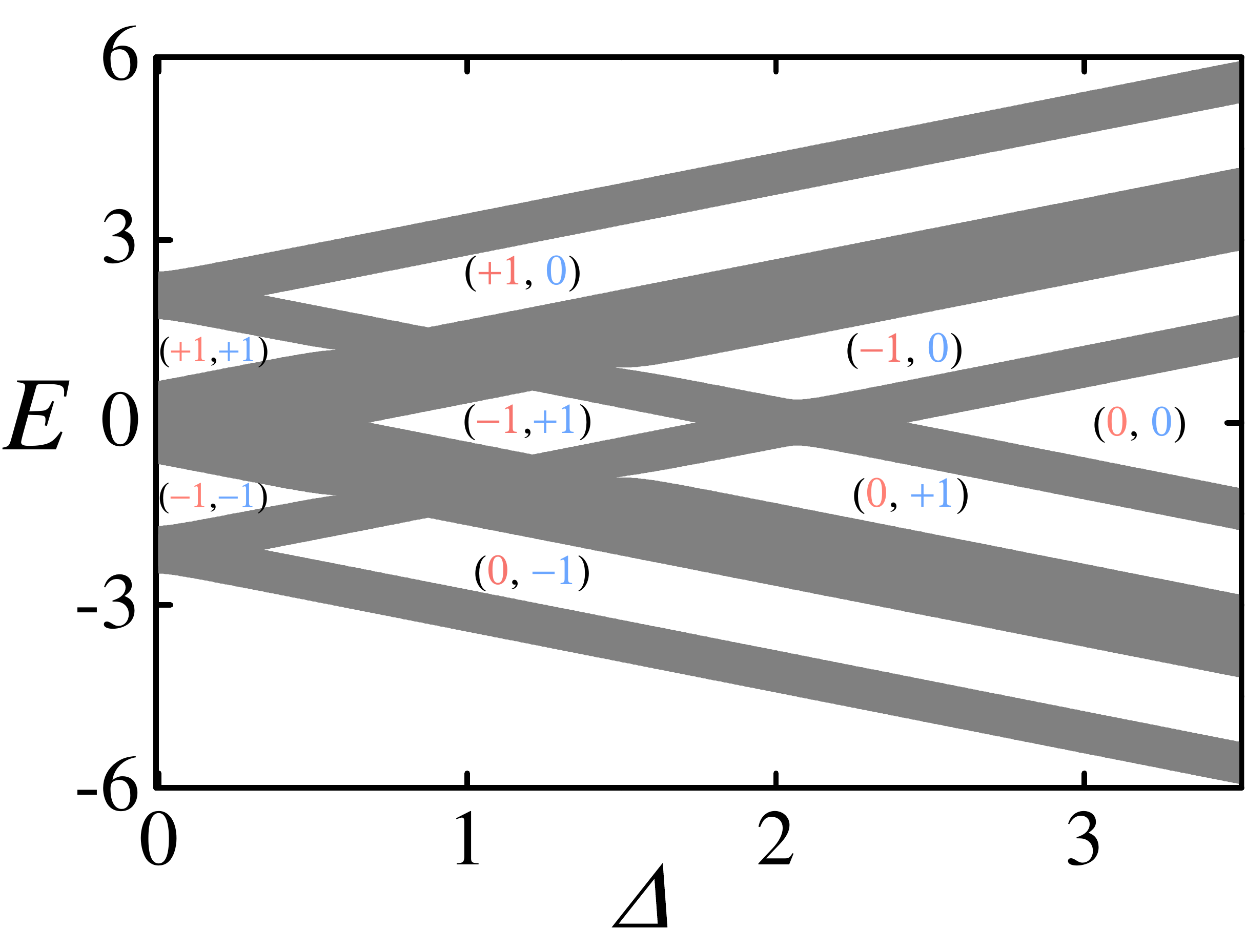}
\caption{$E$-$\Delta$ phase diagram. Here, the
bandgaps are designated by the white regions. The pairs of integers
indicate the Chern numbers of the bandgaps for spin up and down,
distinguishing the topological regimes of the model.}
\label{fig:phasediagram}
\end{figure}

To further show the rich topological phases, we explore the phase diagram
through tuning the synthetic Hamiltonian. Since changing the energy offset
between legs is readily accessible to the experiment, we compute the phase
diagram as a function of the $\Delta$. Figure~\ref{fig:phasediagram}
illustrates the diversity of the topological phases. The bulk is insulating
in the white regions and classified into distinct phases: QH
($C_{\uparrow}=-1, C_{\downarrow}=-1$) or ($C_{\uparrow}=+1,
C_{\downarrow}=+1$), QSH ($C_{\uparrow}=-1, C_{\downarrow}=+1$),
spin-filtered QH ($C_{\uparrow}=\pm 1, C_{\downarrow}=0$) or
($C_{\uparrow}=0, C_{\downarrow}=\pm 1$), and ordinary insulator
($C_{\uparrow}=0, C_{\downarrow}=0$). Take $E_\text{Fermi}=0$ and
$E_\text{Fermi}=-1$ for example. With the increase of $\Delta$ the excitation
of $E_\text{Fermi}=0$ will undergo the regimes of metal, QSH insulator, and
ordinary insulator successively. On the other hand, for the
$E_\text{Fermi}=-1$ the energy offset can turn a QH phase into a metal, then
a spin-filtered QH phase. Therefore, by manipulating $\Delta$ and
$E_\text{Fermi}$ the BCL can host rich topological phases.

One of the key features of our scheme is that the spinless ultracold atoms
are utilized and the synthetic spin-orbit structures are built from the leg's
degree of freedom which is coupled to the Abelian gauge field. Neither
internal states (e.g., hyperfine states) nor a non-Abelian gauge is used in
our setting. This effectively avoids the heating in the Raman process if
spin-orbit coupling relies on the excited states~\cite{LinNature}. Besides
that, since the ``spin" itself is synthesized in real
space~\cite{AtalaNP2014}, our method allows for a real-space-resolved
detection of the topological phases, instead of spin-resolved techniques.
This character will facilitate greatly the direct observations of the
spin-orbit coupled topological phases. On the other hand, due to the
versatility of the optical ladders, we remark that our results are also
applicable to the structures prepared in the artificial
dimensions~\cite{StuhlScience2015,ManciniScience2015,ReyPRL2016,Livi2016,YeNature2017,GadwaySciAdv2017,WangPRL2019,ShinPRL2018,ShinPRL2019}.

Several comments are in order. As an alternative to access the
higher-dimensional lattice Hamiltonians that host topological phases,
the topological pumping in lower dimensional systems provides an
additional avenue towards studying topological states of
matter~\cite{ThoulessPRB1983,ZilberbergPRL2012,LohseNP2016,TakahashiNP2016,LohseNature2018,ZilberbergNature2018}.
The pumping experiments of ultracold atoms have been demonstrated in
optical superlattices~\cite{LohseNP2016,TakahashiNP2016}. Therefore,
by adiabatically
and periodically varying a set of BCLs' parameters, e.g., the
relative phase of the bichromatic lattices, one can drive versatile
quantized transports during each cycle, such as
charge~\cite{ThoulessPRB1983} and spin pump~\cite{FuPRB2006}. The
implementation of pumping in our BCL system will arouse the interest
from the experimental side.

In conclusion, we have proposed a system of bichromatic chiral ladder for
studying the topological bands for ultracold atoms, utilizing the concepts of
synthetic spin-orbit coupling and Zeeman splitting. We have demonstrated that
the quantum spin-Hall phase can emerge within this setup. In the meanwhile,
the bichromatic chiral ladders can produce rich topological phases via tuning
the system parameters. We conclude that the bichromatic chiral ladders hence
constitute a surprisingly simple yet versatile scenario to explore synthetic
topological quantum matter for the ultracold atoms. Our proposal to engineer
topological quantum matter is of direct experimental relevance in ultracold
atoms. And the state-of-the-art experimental
setups~\cite{AtalaNP2014,StuhlScience2015,ManciniScience2015,ShinPRL2018}
render our scheme immediately feasible in ultracold atom experiments. In
addition, one of advantages of ultracold atoms systems is the ability to
control the atom-atom interaction via Feshbach resonances. Taking this work
as a basis, we believe that constructing the interacting models within atomic
chiral ladders can trigger the study on the exotic topological many-body
states~\cite{PRX-1,PRX-2}.

\acknowledgements{\emph{Acknowledgements}: The authors are grateful for
support from National Natural Science Foundation of China (11604231,
11574223); Natural Science Foundation of Jiangsu Province (BK20160303,
BK20150303); Natural Science Foundation of the Jiangsu Higher Education
Institutions of China (16KJB140012).}



\begin{thebibliography}{1}

\bibitem{CooperRMP2019} N. R. Cooper, J. Dalibard, and I. B.
    Spielman, ``Topological bands for ultracold atoms," Rev. Mod. Phys. {\bf91}, 015005 (2019).

\bibitem{ZhuAP2019} D. W. Zhang, Y. Q. Zhu, Y. X. Zhao, H. Yan,
    and S. L. Zhu, ``Topological quantum matter with cold atoms," Adv. Phys. {\bf67}, 253 (2019).

\bibitem{Atala:2013} M. Atala, M. Aidelsburger, J. Barreiro,
    D. Abanin, T. Kitagawa, E. Demler, and I. Bloch, ``Direct
    measurement of the Zak phase in topological
    Bloch bands," Nat. Phys. {\bf9}, 795 (2013).

\bibitem{AidelsburgerNP2015} M. Aidelsburger, M. Lohse, C. Schweizer, M.
    Atala, J. T. Barreiro, S. Nascimb¨¨ne, N. R. Cooper, I. Bloch, and N.
    Goldman, ``Measuring the Chern number of Hofstadter bands with
    ultracold bosonic atoms," Nat. Phys. {\bf11}, 162 (2015).

\bibitem{AidelsburgerPRL2013} M. Aidelsburger, M. Atala, M. Lohse, J.
    T. Barreiro, B. Paredes, and I. Bloch, ``Realization of the
    Hofstadter Hamiltonian with Ultracold Atoms in Optical Lattices,"
    Phys. Rev. Lett. {\bf 111}, 185301 (2013).

\bibitem{MiyakePRL2013} H. Miyake, G. Siviloglou, C. Kennedy, W. Burton, and
    W. Ketterle, ``Realizing the Harper Hamiltonian with Laser-Assisted
    Tunneling in Optical Lattices," Phys. Rev. Lett. {\bf111}, 185302 (2013).


\bibitem{Jotzu:2014} G. Jotzu, M. Messer, R. Desbuquois, M. Lebrat, T.
    Uehlinger, D. Greif and T. Esslinger, ``Experimental realization of the
    topological Haldane model with ultracold fermions," Nature {\bf 515}, 237
    (2014).

\bibitem{goldmanprl2010} N. Goldman, I. Satija, P. Nikolic, A. Bermudez, M.
    A. Martin-Delgado, M. Lewenstein, and I. B. Spielman, ``Realistic
    Time-Reversal Invariant Topological Insulators with Neutral Atoms,"
    Phys. Rev. Lett. {\bf105}, 255302 (2010).

\bibitem{KennedyPRL2013} C. Kennedy, G. Siviloglou, H. Miyake, W.
    Burton, and W. Ketterle, ``Spin-Orbit Coupling and Quantum
    Spin Hall Effect for Neutral Atoms without Spin Flips," Phys.
    Rev. Lett. {\bf111}, 225301 (2013).

\bibitem{Paredes} G. Velasco and B. Paredes, ``Classification of topological
    ladder models," arXiv:1907.11460.

\bibitem{AtalaNP2014} M. Atala, M. Aidelsburger, M. Lohse, J. T.
    Barreiro, B. Paredes, and I. Bloch, ``Observation of chiral currents with
    ultracold atoms in bosonic ladders," Nat. Phys. {\bf10}, 588 (2014).

\bibitem{ManciniScience2015} M. Mancini, G. Pagano, G. Cappellini, L.
    Livi, M. Rider, J. Catani, C. Sias, P. Zoller, M. Inguscio, M. Dalmonte, and L.
    Fallani, ``Observation of chiral edge states with neutral fermions
    in synthetic Hall ribbons," Science {\bf349}, 1510 (2015).

\bibitem{StuhlScience2015} B. Stuhl, H. I.
    Lu, L. Aycock, D. Genkina, and I. Spielman, ``Visualizing edge states with an
    atomic Bose gas in the quantum Hall regime," Science {\bf349}, 1514
    (2015).

\bibitem{ReyPRL2016} M. L. Wall, A. P. Koller, S. Li, X. Zhang, N. R.
    Cooper, J. Ye, and A. M. Rey, ``Synthetic Spin Orbit Coupling in an Optical Lattice Clock,"
    Phys. Rev. Lett. {\bf116}, 035301 (2016).
\bibitem{Livi2016} L. Livi, G. Cappellini, M. Diem, L. Franchi, C.
    Clivati, M. Frittelli, F. Levi, D. Calonico, J. Catani, M.
    Inguscio, and L. Fallani, ``Synthetic Dimensions and Spin-Orbit Coupling with an Optical Clock
    Transition," Phys. Rev. Lett. {\bf117}, 220401 (2016).
\bibitem{YeNature2017} S. Kolkowitz, S. L. Bromley,
    T. Bothwell, M. L. Wall, G. E. Marti, A. P. Koller, X. Zhang, A.
    M. Rey, and J. Ye, ``Spin-orbit-coupled fermions in an optical
    lattice clock," Nature (London) {\bf542}, 66 (2017).

\bibitem{GadwaySciAdv2017} F. A. An, E. J. Meier, and
    B. Gadway, ``Direct observation of chiral currents and
    magnetic reflection in atomic flux lattices," Sci. Adv. {\bf3}, e1602685 (2017).

\bibitem{ShinPRL2018} J. H. Kang, J. H. Han, and Y. Shin,
    ``Realization of a cross-linked chiral ladder with neutral
    fermions in an optical lattice by orbital-momentum coupling,"
    Phys.Rev.Lett. {\bf121}, 150403 (2018).
\bibitem{ShinPRL2019} J. H. Han, J. H. Kang, and Y. Shin, ``Band Gap
    Closing in a Synthetic Hall Tube of Neutral Fermions,"
    Phys.Rev.Lett. {\bf122}, 065303 (2019).

\bibitem{WangPRL2019} H. Cai, J. Liu, J. Wu, Y. He, S. Y. Zhu,
    J. X. Zhang, and D. W. Wang, ``Experimental Observation of
    Momentum-Space Chiral Edge Currents in Room-Temperature
    Atoms," Phys. Rev. Lett. {\bf122}, 023601 (2019).

\bibitem{Huegel:2013} D. H{\"u}gel and B. Paredes, ``Chiral ladders and the
    edges of quantum Hall insulators," Phys.Rev. A {\bf89} 023619 (2014).

\bibitem{RoatiNature2008} G. Roati, C. D'Errico, L. Fallani, M. Fattori, C.
    Fort, M. Zaccanti, G. Modugno, M. Modugno, and M. Inguscio, ``Anderson
    localization of a non-interacting Bose Einstein condensate,"
    Nature {\bf 453}, 895 (2008).

\bibitem{ZilberbergPRL2012} Y. E. Kraus, Y. Lahini, Z. Ringel, M.
    Verbin, and O. Zilberberg, ``Topological States and Adiabatic
    Pumping in Quasicrystals," Phys. Rev. Lett. {\bf109}, 106402
    (2012).
\bibitem{ChenPRL2012} L. J. Lang, X. Cai, and S. Chen, ``Edge States
    and Topological Phases in One-Dimensional Optical Superlattices,"
    Phys. Rev. Lett. {\bf108}, 220401 (2012).
\bibitem{VerbinPRL2013} M. Verbin, O. Zilberberg, Y. E. Kraus, Y.
    Lahini, and Y. Silberberg, ``Observation of Topological Phase
    Transitions in Photonic Quasicrystals," Phys. Rev. Lett. {\bf110},
    076403 (2013).
\bibitem{ZilberbergPRL2013} Y. E. Kraus, Z. Ringel, and O.
    Zilberberg, ``Four-Dimensional Quantum Hall Effect in a
    Two-Dimensional Quasicrystal," Phys. Rev. Lett. {\bf111}, 226401
    (2013).
\bibitem{ZilberbergNPhys2016} Y. E. Kraus and O. Zilberberg,
    ``Quasiperiodicity and topology transcend dimensions," Nat. Phys.
    {\bf12}, 624 (2016).

\bibitem{Harper} P. G. Harper, ``Single Band Motion of Conduction Electrons
    in a Uniform Magnetic Field," Proc. Phys. Soc. A {\bf68}, 874 (1955).
\bibitem{Hofstadter} D. Hofstadter, ``Energy levels and wave functions of
    Bloch electrons in rational and irrational magnetic fields," Phys. Rev. B
    {\bf14}, 2239 (1976).

\bibitem{ShengPRL2011} Y. Yang, Z. Xu, L. Sheng, B. Wang, D. Y. Xing,
    and D. N. Sheng, ``Time-reversal-symmetry-broken quantum
    spin Hall effect," Phys. Rev. Lett. {\bf107}, 066602 (2011).

\bibitem{GoldmanEPL2012} N. Goldman, W. Beugeling, and C. Smith,
    ``Topological phase transitions between chiral and helical spin with
    spin-orbit coupling and a magnetic field-tight binding,"
    Europhys. Lett. {\bf97}, 23003 (2012).
\bibitem{BeugelingPRB2012} W. Beugeling, N. Goldman, and C. M. Smith,
    ``Topological phases in a two-dimensional lattice: Magnetic field versus
    spin-orbit coupling," Phys. Rev. B {\bf86}, 075118 (2012).

\bibitem{LinNature} Y. J. Lin, K. Jim\'{e}nez-Garc\'{i}a, and I. B. Spielman,
    ``Spin-orbit-coupled Bose-Einstein condensates," Nature {\bf471}, 83
    (2011).

\bibitem{ThoulessPRB1983} D. J. Thouless, ``Quantization of particle
    transport,", Phys. Rev. B {\bf27}, 6083 (1983).

\bibitem{LohseNP2016}  M. Lohse, C. Schweizer, O. Zilberberg, M.
    Aidelsburger, and I. Bloch, ``A Thouless Quantum Pump with
    Ultracold Bosonic Atoms in an Optical Superlattice," Nat. Phy. {\bf12}, 350
    (2016).
\bibitem{TakahashiNP2016}S. Nakajima, T. Tomita, S. Taie, T. Ichinose, H.
    Ozawa, L. Wang, M. Troyer, and Y. Takahashi, ``Topological Thouless
    pumping of ultracold fermions," Nat. Phys. {\bf12}, 296 (2016).

\bibitem{LohseNature2018} M. Lohse, C. Schweizer, H. M. Price, O.
    Zilberberg, and I. Bloch, ``Exploring 4D quantum Hall physics with a 2D topological
    charge pump," Nature {\bf553}, 55 (2018).

\bibitem{ZilberbergNature2018} O. Zilberberg, S. Huang, J.
    Guglielmon, M. Wang, K. P. Chen, Y. E. Kraus, and M. C. Rechtsman, ``Photonic topological boundary
    pumping as a probe of 4D quantum Hall physics," Nature {\bf553},
    59 (2018).

\bibitem{FuPRB2006} L. Fu and C. Kane, ``Time reversal polarization
    and a $Z_2$ adiabatic spin pump," Phys. Rev. B {\bf74},
    195312 (2006).

\bibitem{PRX-1} M. C. Strinati, E. Cornfeld, D. Rossini,
    S. Barbarino, M. Dalmonte, R. Fazio, E. Sela, and
    L. Mazza, ``Laughlin-like States in Bosonic and Fermionic Atomic
    Synthetic Ladders," Phys. Rev. X {\bf7}, 021033 (2017).

\bibitem{PRX-2} J. J\"{u}nemann, A. Piga, S. J. Ran, M. Lewenstein, M.
    Rizzi, and A. Bermudez, ``Exploring Interacting
    Topological Insulators with Ultracold
    Atoms: The Synthetic Creutz-Hubbard Model," Phys. Rev. X {\bf7}, 031057
    (2017).





%
%
%
%
%
%
%
%
%
%
%
%
%
%
%

\end{thebibliography}
\end{document}